\documentclass[superscriptaddress,nobibnotes,amsmath,amssymb,notitlepage,twocolumn,pra]{revtex4-2}

\usepackage{titlesec}
\titleformat{\section}
  {\centering\normalfont\fontsize{12}{15}\bfseries}{\thesection}{1em}{}
\makeatletter
\def\@hangfrom@section#1#2#3{\@hangfrom{#1#2}#3}
\def\@hangfroms@section#1#2{#1#2}
\makeatother

\usepackage{bm,braket}
\usepackage[toc,page]{appendix}
\usepackage{comment}
\usepackage{dcolumn}
\usepackage{ulem}
\usepackage{amsfonts}

\usepackage{graphicx,color,hyperref}
\usepackage[caption=false]{subfig}
\hypersetup{colorlinks=true, linkcolor=blue, citecolor=blue, urlcolor=blue} 

\newcommand{\beq}[1]{\begin{equation}\label{#1}}
\newcommand{\eep}{\;.\end{equation}}
\newcommand{\eec}{\;,\end{equation}}
\newcommand{\eeq}{\end{equation}}

\makeatletter
\newcommand{\fmarki}{*}
\newcommand{\fmarkii}{\ensuremath{\dagger}}
\newcommand{\fmarkiii}{\ensuremath{\ddagger}}
\newcommand{\fmarkiv}{\ensuremath{\mathsection}}
\newcommand{\fmarkv}{\ensuremath{\mathparagraph}}
\newcommand{\fmarkvi}{\ensuremath{\|}}
\newcommand{\fmarkvii}{**}
\newcommand{\fmarkviii}{\ensuremath{\dagger\dagger}}
\newcommand{\fmarkix}{\ensuremath{\ddagger\ddagger}}
                
\def\@fnsymbol#1{{\ifcase#1\or \fmarki\or \fmarkii\or \fmarkiii\or \fmarkiv\or \fmarkv\or \fmarkvi\or \fmarkvii\or \fmarkviii\or \fmarkix \else\@ctrerr\fi}}
\makeatother

\renewcommand{\fmarki}{$*$}
\renewcommand{\fmarkii}{$*$}
\renewcommand{\fmarkiii}{$*$}
\renewcommand{\fmarkiv}{$*$}
\renewcommand{\fmarkv}{$*$}
\renewcommand{\fmarkix}{$*$}

\DeclareMathAlphabet{\mathcal}{OMS}{cmsy}{m}{n} 

\usepackage{amsmath}
\usepackage{amssymb}
\usepackage{xcolor}
\usepackage{bbm}
\usepackage{physics}
\usepackage{float}
\usepackage{graphicx}
\usepackage{dcolumn} 
\usepackage{bm} 
\usepackage{siunitx}
\usepackage{enumitem}  

\makeatletter
\renewcommand*{\fnum@figure}{{\normalfont\bfseries \figurename~\thefigure}}
\makeatother

\allowdisplaybreaks

\definecolor{orange}{rgb}{1,0.5,0}

\newcommand{\supl}[1]{Supplement S.#1}

\graphicspath{{./img/}}

\DeclareMathAlphabet{\mathcal}{OMS}{cmsy}{m}{n} 

\makeatletter
\newcommand{\specificthanks}[1]{\@fnsymbol{#1}}
\makeatother

\begin{document}

\raggedbottom

\preprint{APS/123-QED}

\title{Enhancement of exciton radius near a band-gap closing through quantum geometry}

\author{Jin-Hyung~Choi}
\thanks{}
\affiliation{Department of Physics, Hanyang University, Seoul 04763, Republic of Korea}
\affiliation{Department of Physics, Ajou University, Suwon 16499, Republic of Korea}

\author{Sang-Hoon~Han}
\thanks{}
\affiliation{Department of Physics, Ajou University, Suwon 16499, Republic of Korea}

\author{Young-Kwon~Han}
\thanks{}
\affiliation{Department of Physics, Hanyang University, Seoul 04763, Republic of Korea}

\author{Sun-Woo~Kim}
\thanks{}
\email{sunwookim@hanyang.ac.kr}
\affiliation{Department of Physics, Hanyang University, Seoul 04763, Republic of Korea}

\author{Jun-Won~Rhim}
\email{jwrhim@ajou.ac.kr}
\affiliation{Department of Physics, Ajou University, Suwon 16499, Republic of Korea}

\author{Joshua J. P. Thompson}
\email{jjt56@cam.ac.uk}
\affiliation{Department of Materials Science and Metallurgy, University of Cambridge,
27 Charles Babbage Road, Cambridge CB3 0FS, United Kingdom}

\date{\today}

\begin{abstract}
Exciton engineering traditionally focuses on modifying semiclassical material properties, such as the effective mass and dielectric screening, while largely overlooking the quantum geometry of the underlying electron and hole Bloch states. This approximation is adequate for many materials but breaks down near a topological band-gap closing, where the quantum metric around the band extrema becomes strongly enhanced. In this regime, Bloch states at different momenta become less similar, reducing the projected electron--hole Coulomb matrix elements and consequently weakening exciton binding. We demonstrate this mechanism in a spin--orbit-coupled Lieb-lattice model tuned toward a topological phase transition. The suppressed Coulomb matrix elements narrow the exciton wavefunction in momentum space, leading to an enlarged exciton radius in real space. This increase in exciton size produces an experimentally accessible enhancement of the weak-field diamagnetic response. Our results show that quantum geometry can fundamentally reshape exciton properties near a topological phase transition, revealing a previously underexplored route for engineering excitonic states.
\end{abstract}

\maketitle

\section*{Introduction}
Excitons are fundamental excitations governing the optical response of semiconductors and insulators, with their energies and wavefunctions determined jointly by the electronic band structure and the electron--hole Coulomb interaction. In reduced-dimensional or weakly screened materials, excitons dominate the optical spectra and energy transport, making their internal quantum structure critical in determining optoelectronic functionality \cite{Wannier1937,Mott1938,HankeSham1980,Strinati1988,Onida2002,Rohlfing2000,Qiu2013,Berkelbach2013,wang2018colloquium,perea2022exciton}. Within the conventional Wannier--Mott picture, the exciton is described by a slowly varying relative-motion wavefunction determined primarily by the band dispersion near the gap and the screened Coulomb interaction. Effective-mass and low-order $k\cdot p$ approaches restrict the problem to momenta near the band edge, as in treatments of excitons in layered perovskites and transition-metal dichalcogenide monolayers \cite{Thompson2024PhononBottleneck,Berghaeuser2014Analytical}. In their simplest envelope-function implementation, however, these approaches neglect the momentum dependence of the cell-periodic Bloch functions over the range relevant to exciton formation.

This approximation neglects a geometric contribution to the band-projected electron--hole interaction. In a band-projected description, the Coulomb matrix elements contain overlaps between cell-periodic Bloch wavefunctions at different momenta, which act as scattering form factors for the Coulomb coupling. The commonly adopted unity-overlap approximation replaces these form factors by unity, effectively assuming that the Bloch frame \cite{PanatiPisante2013BlochFrame,Cances2017BlochFrame} remains constant over the momentum scale relevant to exciton formation. This approximation is valid when neighbouring Bloch states vary only weakly in Hilbert space, but breaks down once the Bloch frame acquires strong momentum dependence.

Quantum geometry provides a natural framework for describing this breakdown. The Berry curvature and quantum metric characterize the geometric phase structure and quantum distance between Bloch states in momentum space, respectively \cite{Berry1984,Provost1980,Xiao2010,Bouhon2023QuantumGeometry,Yu2025QuantumGeometry,rhim2020quantum,rhim2021singular}. Previous studies have shown that Berry curvature and band topology influence exciton spectra, optical selection rules, and other geometric properties \cite{Srivastava2015,Zhou2015,Zhang2018,Cao2018,Tang2024Inheritance,QiuWu2025,Lozano2025}. More recently, interaction-induced exciton topology, Berry-phase descriptions of exciton polarization, and geometric effects on exciton transport have further demonstrated that excitons can inherit or acquire nontrivial geometric structure from both the underlying electronic states and the electron--hole interaction \cite{Davenport2024Interaction,Davenport2026Berryology,Jankowski2025OrganicExcitons,Thompson2025ExcitonTransport}. Despite these advances, it remains unclear how the quantum distance between Bloch states modifies the projected Coulomb interaction and ultimately controls the spatial extent of an exciton.

In this work, we address this question by developing a gauge-invariant formulation of the exciton radius in a band-projected basis and applying it to a spin--orbit-coupled Lieb-lattice model tuned towards a band-topology transition \cite{Weeks2010,Goldman2011,Leykam2018,rhim2026topological}. We show that quantum geometry substantially enlarges the exciton radius through geometric suppression of the projected electron--hole Coulomb interaction. Near a band-gap closing, the rapid momentum variation of the Bloch frame enhances the quantum metric and reduces the scattering form factors entering the projected Coulomb matrix elements. The resulting suppression of finite-momentum Coulomb scattering localizes the lowest-energy exciton wavefunction around the band-gap minimum in momentum space and, through the Fourier relation, enlarges its real-space radius.

To quantify this effect, we evaluate the electron--hole relative separation using the band-projected position operator, whose matrix elements contain Berry-connection contributions associated with the underlying Bloch states \cite{Resta1998PositionOperator,Ahn2022RiemannianOptics}. This formulation naturally yields a gauge-invariant exciton radius that incorporates both the excitonic wavefunction and the geometry of the Bloch frame. Comparison with the conventional unity-overlap approximation demonstrates a pronounced enhancement of the exciton radius near the band-topology transition. We further show that this enhancement produces an increased weak-field diamagnetic response, establishing the exciton radius as an experimentally accessible probe of quantum geometry \cite{Stier2016Diamagnetic,Goryca2019HighField}.

\begin{figure} [!t]
    \centering
    \includegraphics[width=\linewidth]{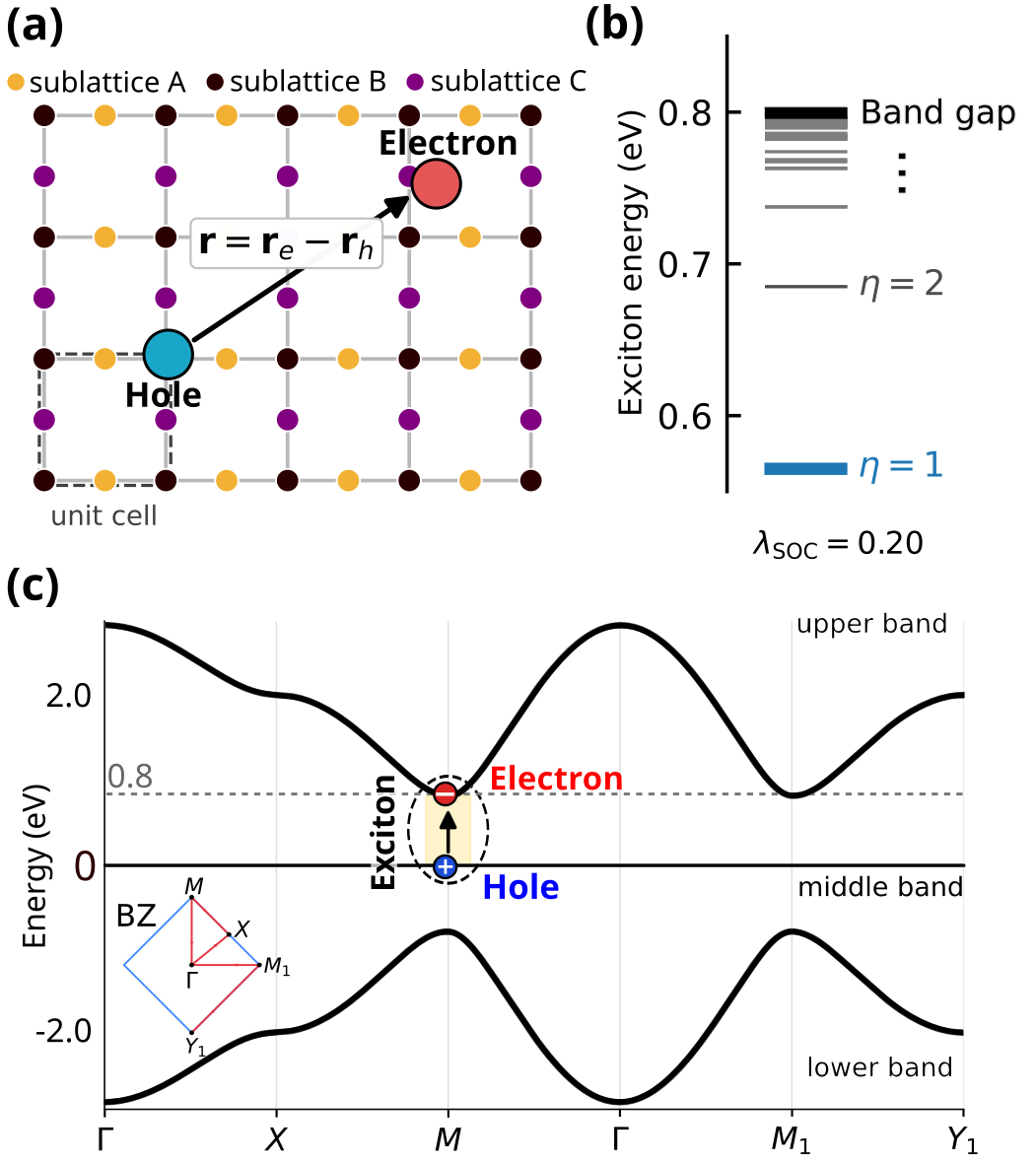} 
\caption{{\bf Overview of the Lieb-lattice exciton.} 
(a) Lieb lattice with three sublattices per unit cell and an exciton represented by the in-plane electron--hole relative coordinate $\mathbf r=\mathbf r_e-\mathbf r_h$.
(b) Energies of lowest exciton states at
$\lambda_{\rm SOC}=0.20$. The black line marks the minimum direct
single-particle band gap.
(c) Single-particle band structure along the high-symmetry path shown in the inset.
The exciton is formed between the upper and middle bands.
The single-particle direct optical gap is approximately 0.8 eV in this case, and the exciton levels lie below this energy owing to the attractive electron--hole interaction.}
    \label{fig:fig1}
\end{figure}

\section*{Results}

\subsection*{Band-projected exciton and gauge-covariant radius}
\label{sec:projected_exciton_radius}

We consider a zero-centre-of-mass exciton formed between an isolated conduction band $c$ and valence band $v$. We focus throughout on the lowest-energy exciton. The corresponding exciton state is written as
\begin{equation}
|\Phi\rangle
=
\sum_{\mathbf k}
\varphi(\mathbf k)
\hat c^\dagger_{c\mathbf k}
\hat c_{v\mathbf k}
|\mathrm{GS}\rangle ,
\label{eq:exciton_state}
\end{equation}
where $\varphi(\mathbf k)$ is the normalized momentum-space wavefunction of the lowest-energy exciton and $|\mathrm{GS}\rangle$ is the many-electron ground state.
The energy $E$ and exciton wavefunction $\varphi(\mathbf k)$ are obtained as the lowest eigenpair of the band-projected Wannier equation, equivalently the direct electron--hole Bethe--Salpeter equation within the selected two-band subspace \cite{Wannier1937,HankeSham1980,Strinati1988,Rohlfing2000,Onida2002},
\begin{equation}
\left[
\varepsilon_c(\mathbf k)
-
\varepsilon_v(\mathbf k)
\right]
\varphi(\mathbf k)
+
\sum_{\mathbf k'}
V_0(\mathbf k-\mathbf k')
\Lambda_{\mathbf k\mathbf k'}
\varphi(\mathbf k')
=
E\,
\varphi(\mathbf k).
\label{eq:projected_wannier}
\end{equation}
Here, $\varepsilon_c(\mathbf k)-\varepsilon_v(\mathbf k)$ is the noninteracting electron--hole excitation energy, $E$ is the energy of the exciton, and $V_0(\mathbf k-\mathbf k')<0$ denotes the screened attractive Coulomb potential. Projection onto the conduction and valence bands introduces the scattering form factor $\Lambda_{\mathbf k\mathbf k'}=\langle u_{c\mathbf k}|u_{c\mathbf k'}\rangle\langle u_{v\mathbf k'}|u_{v\mathbf k}\rangle$, where $|u_{n\mathbf k}\rangle$ is the cell-periodic Bloch state of band $n$. The projected interaction matrix element is therefore
\begin{equation}
V_{\mathbf k\mathbf k'}
=
V_0(\mathbf k-\mathbf k')
\Lambda_{\mathbf k\mathbf k'} .
\label{eq:projected_interaction}
\end{equation}
The magnitude of the form factor is controlled by the quantum distances \cite{Srivastava2015,Zhou2015,Yu2025QuantumGeometry,rhim2020quantum,rhim2021singular}, $d_{\mathrm{HS}}^{eh}(\mathbf k,\mathbf k') = \sqrt{1-|\Lambda_{kk'}|^2}$, between the conduction- and valence-band Bloch states connected by the Coulomb interaction. It equals unity at $\mathbf k=\mathbf k'$ but can fall well below unity when either set of Bloch states varies rapidly over the relevant momentum range.

To isolate the effect of the momentum-dependent Bloch frame, we compare the calculation retaining the momentum-dependent Bloch frame with a unity-overlap reference. In the latter, the scattering form factor is replaced by $\Lambda_{\mathbf k\mathbf k'}=1$, while the single-particle dispersions and screened Coulomb potential are kept identical. The reference Wannier equation consequently retains the same Coulomb-induced coupling between different relative momenta but removes its suppression by Bloch-state overlaps. The distinction between the two calculations is therefore not the presence or absence of the Coulomb interaction, but the strength and momentum dependence of its band-projected matrix elements.

The internal radius of an exciton is determined by the electron--hole relative-coordinate operator. In a band-projected basis, this operator cannot be represented by an ordinary momentum derivative alone because the electron--hole Bloch frame also varies with crystal momentum\cite{Blount1962,Resta1998PositionOperator,Ahn2022RiemannianOptics,Davenport2026Berryology}. For the band-projected exciton considered here, the momentum-dependent electron--hole product Bloch frame is written as $|U_{\mathbf k}\rangle = |u_{c\mathbf k}\rangle \otimes |u^*_{v\mathbf k}\rangle$\cite{Jankowski2025OrganicExcitons,Thompson2025ExcitonTransport}.
The Berry connection associated with this product frame is
\begin{equation}
\mathcal A^{eh}_{\mu}(\mathbf k) = i\langle U_{\mathbf k}| \partial_{\mu} U_{\mathbf k}\rangle = \mathcal A^{c}_{\mu}(\mathbf k) - \mathcal A^{v}_{\mu}(\mathbf k),
\label{eq:electron_hole_connection}
\end{equation}
where $\mathcal A^{n}_{\mu}(\mathbf k) = i\langle u_{n\mathbf k}| \partial_{\mu} u_{n\mathbf k}\rangle $ is the Berry connection of band $n$ and $\partial_\mu = \partial/\partial k_\mu$. The projected relative-coordinate operator is then
\begin{equation}
\hat r_\mu
=
iD_\mu,
\label{eq:covariant_relative_position}
\end{equation}
where $D_\mu
=
\partial_{\mu}
-
i\mathcal A^{eh}_{\mu}(\mathbf k)$.
The covariant derivative acts on the exciton section formed jointly by the exciton wavefunction $\varphi(\mathbf k)$ and the
momentum-dependent electron--hole Bloch frame. Under a change of local frame, the two components transform oppositely. Consequently, the covariant derivative of the section is independent of the choice of local frame; in the chosen frame, it is represented by $D_\mu\varphi$, and the resulting relative-coordinate matrix elements are gauge invariant.

We define the squared exciton radius as the variance of the relative-coordinate operator,
$\xi^2
=
\sum_{\mu=x,y}
\left[
\langle \hat r_\mu^2\rangle
-
\langle \hat r_\mu\rangle^2
\right]$.
Using Eq.~\eqref{eq:covariant_relative_position} together with the
Brillouin-zone sewing conditions, this can be expressed as
\begin{equation}
\xi^2
=
\sum_{\mu=x,y}
\left[
\sum_{\mathbf{k}}|D_\mu\varphi(\mathbf{k})|^2
-
\left|
\sum_{\mathbf{k}}
\varphi^{*}(\mathbf{k})
D_\mu
\varphi(\mathbf{k})
\right|^2
\right],
\label{eq:exciton_radius_covariant}
\end{equation}
where the inner products are evaluated over crystal momentum. This expression is gauge invariant and measures the real-space electron--hole separation within the projected exciton state. It also shows that the radius cannot, in general, be inferred solely from the scalar width of $|\varphi(\mathbf k)|^2$.

Although the separate derivative, Berry-connection, and intracell-embedding contributions depend on the tight-binding Bloch convention, the physical projected interaction matrix elements and the covariant exciton radius are invariant under a momentum-dependent change of orbital basis when all quantities are transformed consistently. We use the convention in which the intracell orbital positions are included in the tight-binding Bloch phases. The corresponding Brillouin-zone sewing conditions and position-operator derivation are given in \supl \textrm{I}.

For a consistent unity-overlap reference, the momentum dependence of the electron--hole product Bloch frame is removed from both the projected interaction and the radius evaluation. We denote the lowest-energy solution of the reference Wannier equation by $E^{(0)}$ and $\varphi^{(0)}(\mathbf k)$, and its radius by $\xi^{(0)}$. The reference wavefunction is obtained by setting $\Lambda_{\mathbf k\mathbf k'}=1$ while retaining the same single-particle dispersions and screened Coulomb potential, and its radius is evaluated using the ordinary derivative,
$
\bigl(\xi^{(0)}\bigr)^2
=
\sum_{\mu}
\left[
\sum_{\mathbf{k}}|\partial_\mu\varphi^{(0)}(\mathbf{k})|^2
-
\left|
\sum_{\mathbf{k}}
\varphi^{(0)*}(\mathbf{k})
\partial_\mu
\varphi^{(0)}(\mathbf{k})
\right|^2
\right]$.
The comparison between $(E,\varphi,\xi)$ and $(E^{(0)},\varphi^{(0)},\xi^{(0)})$ therefore quantifies the total Bloch-frame contribution within the band-projected description.

Within the weak-field effective-mass approximation, the squared radius determines the leading orbital response to a perpendicular magnetic field. The quadratic diamagnetic energy shift is
\begin{equation}
\Delta E_{\mathrm{dia}}(B)
=
\gamma_{\mathrm{dia}} B^2,
\label{eq:diamagnetic_shift}
\end{equation}
where
$\gamma_{\mathrm{dia}}=e^2\xi^2/(8m_{\mathrm r})$
and $m_{\mathrm r}$ is the electron--hole reduced mass \cite{Stier2016Diamagnetic,Goryca2019HighField}.
For the unity-overlap reference, the corresponding quantities are $\Delta E_{\mathrm{dia}}^{(0)}(B)
=
\gamma_{\mathrm{dia}}^{(0)}B^2$, and $\gamma_{\mathrm{dia}}^{(0)}
=
e^2\bigl(\xi^{(0)}\bigr)^2/(8m_{\mathrm r})$.
Because both calculations retain the same single-particle dispersions, the same reduced mass is used in evaluating their diamagnetic coefficients. Comparison with the unity-overlap reference therefore quantifies the net effect of the momentum-dependent Bloch frame on the exciton radius and its diamagnetic response. The diamagnetic coefficient consequently provides an experimentally accessible signature of this geometric modification. 

\subsection*{Spin--orbit-coupled Lieb lattice}
\label{sec:lieb_lattice_excitons}

We apply the band-projected formulation to a spin--orbit-coupled Lieb lattice, which provides a minimal setting in which a band gap and the momentum dependence of the Bloch states can be tuned systematically \cite{Weeks2010,Goldman2011,Leykam2018,rhim2026topological}. As illustrated in Fig.~\ref{fig:fig1}(a), the lattice contains three sublattice orbitals, denoted $A$, $B$, and $C$, in each unit cell. The exciton consists of an electron and a hole whose internal coordinate is the in-plane relative displacement $\mathbf r=\mathbf r_e-\mathbf r_h$. The nearest-neighbour hopping produces a middle band between two dispersive bands, while spin--orbit coupling opens gaps between them. In the following, $\lambda_{\rm SOC}$ is used to tune the gap between the upper and middle bands, from which the electron and hole states are constructed, respectively.

Figure~\ref{fig:fig1}(c) shows the single-particle band structure for $\lambda_{\rm SOC}=0.20$. The middle band lies at zero energy, whereas the upper and lower bands are dispersive and separated from it by the spin--orbit-induced gaps. The exciton considered here is formed by promoting an electron from the middle band to the upper band at the same crystal momentum, consistent with the zero-centre-of-mass condition in Eq.~\eqref{eq:exciton_state}. The minimum direct electron--hole excitation energy occurs at the $M$ point and the symmetry-related band-gap minima.
For the parameters used in Fig.~\ref{fig:fig1}, the corresponding single-particle gap is approximately $0.8\,\mathrm{eV}$. This band-gap minimum defines the momentum-space region relevant to the exciton state examined below.

The lowest eigenvalue of the projected Wannier equation lies below the minimum direct electron--hole excitation energy, demonstrating that the projected attractive interaction supports a bound exciton in the upper--middle-band subspace. Figure~\ref{fig:fig1}(b) shows its energy $E$ at $\lambda_{\rm SOC}=0.20$, together with the minimum direct band gap. The lowest-energy state is the exciton considered throughout the remainder of this work.


\begin{figure} [!t]
    \centering
    \includegraphics[width=\linewidth]{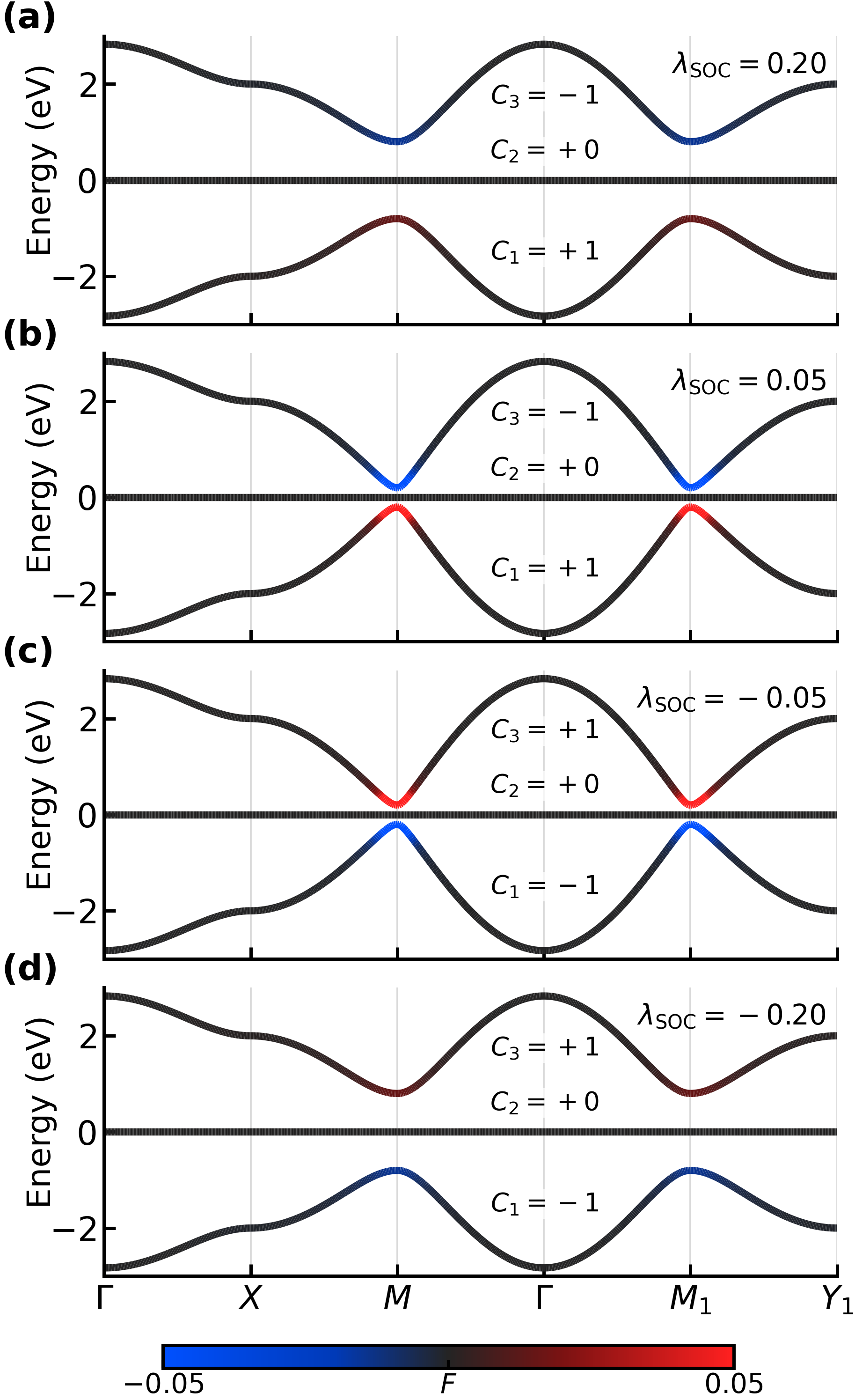} 
\caption{
{\bf SOC-tuned band geometry of the Lieb-lattice model.}
(a-d) Band structures for $\lambda_{\rm SOC}=0.20$, $0.05$, $-0.05$, and $-0.20$.
The colour scale shows the Berry curvature along the high-symmetry path, and the labels indicate the Chern numbers of the corresponding bands.
As $|\lambda_{\rm SOC}|$ decreases, the bands become more Dirac-like near the band-gap minimum and the Berry curvature becomes more strongly localized there.
Although the Berry-curvature distribution changes substantially, the band Chern numbers remain fixed until the gap closes at $\lambda_{\rm SOC}=0$.
}
    \label{fig:fig2}
\end{figure}

\subsection*{Band topology near the gap closing}
\label{sec:band_topology}

We examine how the single-particle gap and band topology evolve as the spin--orbit coupling is tuned toward zero. Figure~\ref{fig:fig2} shows the band dispersions for $\lambda_{\rm SOC}=0.20$, $0.05$, $-0.05$, and $-0.20$, with the colour indicating the band-resolved Berry curvature $\mathcal F_n(\mathbf k)$. The corresponding Chern number of band $n$ is
\begin{equation}
C_n
=
\frac{1}{2\pi}
\int_{\mathrm{BZ}}
d^2k\,
\mathcal F_n(\mathbf k).
\label{eq:chern_number}
\end{equation}

For positive spin--orbit coupling, the three bands carry $(C_1,C_2,C_3)=(+1,0,-1)$, as shown in Figs.~\ref{fig:fig2}(a,b). Decreasing $\lambda_{\rm SOC}$ from $0.20$ to $0.05$ reduces the direct gap between the upper and middle bands at the $M$ point and its symmetry-related counterpart $M_1$, while leaving the Chern numbers unchanged. The minimum direct gap decreases from approximately $0.8\,\mathrm{eV}$ at $\lambda_{\rm SOC}=0.20$ to $0.2\,\mathrm{eV}$ at $\lambda_{\rm SOC}=0.05$. The same evolution occurs for negative spin--orbit coupling, as shown in Figs.~\ref{fig:fig2}(c,d), with the magnitude of the gap controlled by $|\lambda_{\rm SOC}|$. The explicit Bloch Hamiltonian, reciprocal-space sewing matrices, and Chern-number conventions are summarized in \supl \textrm{II}.

The gap closes at $\lambda_{\rm SOC}=0$. Across this closing, the Chern numbers of the upper and lower bands reverse sign,
\begin{equation}
(C_1,C_2,C_3)
=
\begin{cases}
(+1,0,-1), & \lambda_{\rm SOC}>0,\\
(-1,0,+1), & \lambda_{\rm SOC}<0,
\end{cases}
\label{eq:lieb_chern_numbers}
\end{equation}
whereas the middle band remains topologically trivial. Thus, $\lambda_{\rm SOC}=0$ separates two gapped phases with opposite Chern-band chiralities and constitutes the band topology transition considered below. The exciton is evaluated on both sides of this transition, where the upper and middle bands remain isolated.

\begin{figure*}[!t]
\centering
\includegraphics[width=0.8\textwidth]{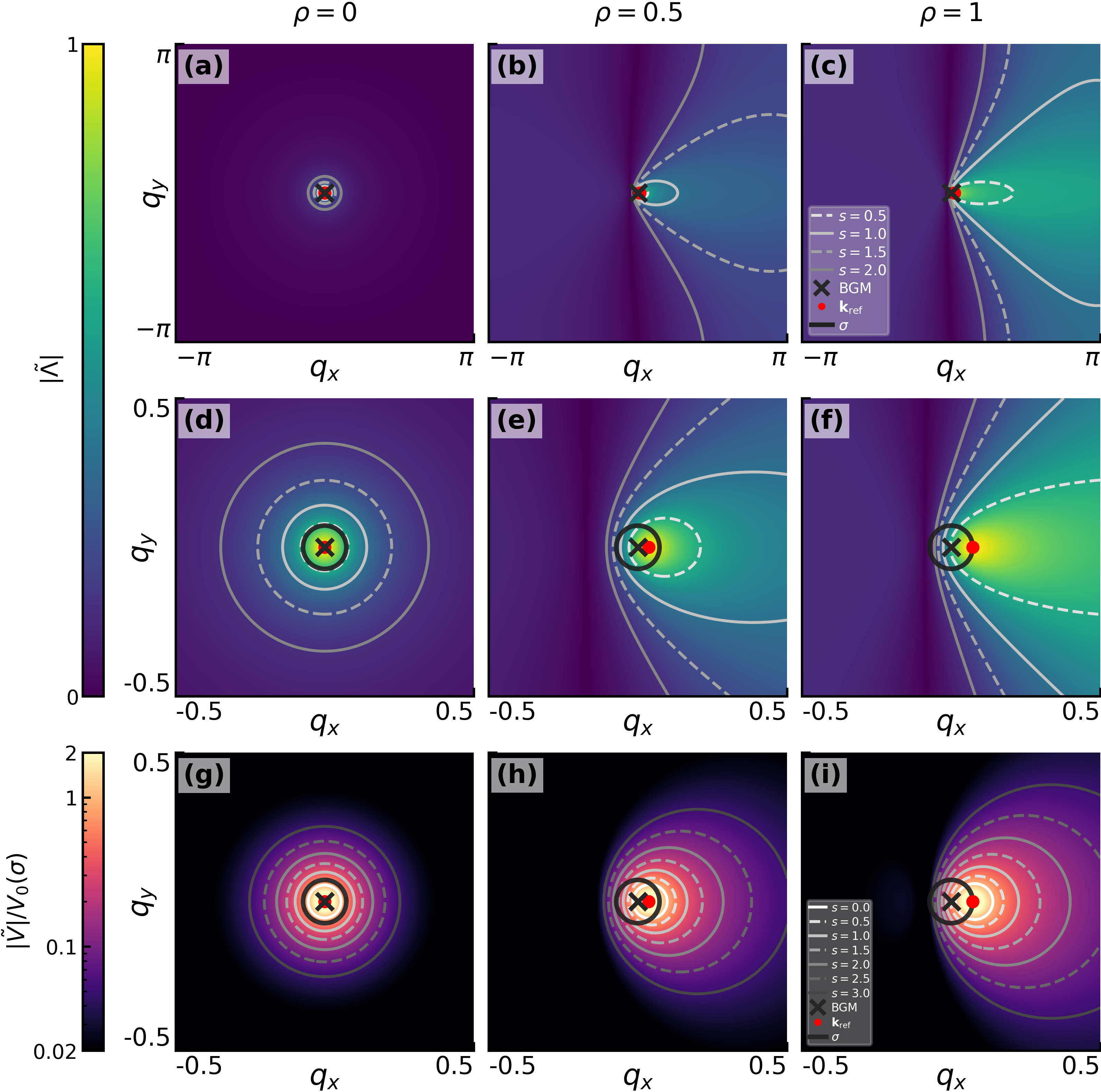}
\caption{
\textbf{Form-factor suppression of the projected Coulomb interaction.}
Panels (a)--(f) show the electron--hole form-factor magnitude $|\widetilde{\Lambda}_{\mathbf q_{\rm ref}\mathbf q}|$, and panels (g)--(i) show the corresponding projected interaction magnitude $|\widetilde V_{\mathbf q_{\rm ref}\mathbf q}|/|V_0(\sigma)|$. Here,
$\mathbf q=\mathbf k-\mathbf M$,
$\mathbf q_{\rm ref}=\mathbf k_{\rm ref}-\mathbf M
=\rho\,\sigma\,(\cos\theta,\sin\theta)$, and
$\widetilde V_{\mathbf q_{\rm ref}\mathbf q}
=
V_0(\mathbf q_{\rm ref}-\mathbf q)
\widetilde{\Lambda}_{\mathbf q_{\rm ref}\mathbf q}$. The results are shown for $\theta=0$ and $\rho=0$, $0.5$, and $1.0$ from left to right. Panels (a)--(c) show the full momentum window, whereas panels (d)--(f) and (g)--(i) show the corresponding regions around the band-gap minimum $\mathbf M$. The contours mark $|\widetilde{\Lambda}_{\mathbf q_{\rm ref}\mathbf q}|=e^{-s}$ in panels (a)--(f) and $|\widetilde V_{\mathbf q_{\rm ref}\mathbf q}|/|V_0(\sigma)|=e^{-s}$ in panels (g)--(i), for the indicated values of $s$. The black circle has radius $\sigma$, where $\sigma$ is the momentum-space standard deviation of $|\varphi(\mathbf k)|^2$ about $\mathbf M$; the cross marks $\mathbf M$, and the red point marks $\mathbf q_{\rm ref}$. For $\mathbf q_{\rm ref}\neq\mathbf 0$, the form factor suppresses coupling away from the reference momentum and preferentially retains larger matrix elements along approximately the same radial direction from $\mathbf M$, producing an anisotropic projected interaction.
}
\label{fig:fig3}
\end{figure*}

The local distribution of Berry curvature changes strongly even within each gapped phase. At $|\lambda_{\rm SOC}|=0.20$, the curvature is spread over a relatively broad momentum range. As the gap decreases, it becomes increasingly concentrated around $M$ and $M_1$, where the relevant band separations are smallest. The sign of the curvature reverses when $\lambda_{\rm SOC}$ changes sign, consistently with the reversal of the band Chern numbers. Consequently, tuning toward the gap closing produces a pronounced reorganization of the local Bloch-state texture while the Chern numbers remain fixed on each side of the transition.

The Berry-curvature concentration signals that the momentum dependence of the Bloch states becomes particularly strong near the band-gap minima. Berry curvature alone, however, does not determine the quantum-distance reduction of finite-momentum overlaps. We therefore examine next the scattering form factors entering the projected Coulomb matrix elements, which provide a direct measure of how this rapidly varying Bloch structure modifies the exciton problem.

\subsection*{Quantum-distance suppression of the projected Coulomb coupling} \label{sec:projected_coulomb_suppression}
We now examine how the rapid momentum variation of the Bloch frame near the band-gap closing modifies the projected Coulomb matrix elements. For two normalized electron--hole product Bloch frames at arbitrary crystal momenta, their finite separation in projective Hilbert space can be quantified by the squared Hilbert--Schmidt quantum distance \cite{Hwang2021FlatBandLandau,Yu2025QuantumGeometry},
\begin{equation}
\left[ d_{\mathrm{HS}}^{eh}(\mathbf k,\mathbf k') \right]^2 = 1- \left| \langle U_{\mathbf k}|U_{\mathbf k'}\rangle \right|^2 = 1- \left| \Lambda_{\mathbf k\mathbf k'} \right|^2 .
\label{eq:electron_hole_hs_distance}
\end{equation}
Unlike the local quantum metric, this expression directly characterizes the separation between Bloch frames at arbitrary momenta. In the infinitesimal limit, its leading second-order variation is determined by the quantum metric \cite{Provost1980}. More generally, the quantum metric defines the path length associated with a trajectory through momentum space and thereby describes how rapidly the Bloch frame varies along that trajectory. At large momentum separations, this quantum-metric path length need not coincide quantitatively with the Hilbert--Schmidt distance between the endpoints, although both reflect the same underlying Bloch-state geometry. We therefore use Eq.~\eqref{eq:electron_hole_hs_distance} to quantify finite separations and the quantum metric to interpret their momentum dependence.

For convenience, we use momentum coordinates centred at the band-gap minimum, defined by $\mathbf q=\mathbf k-\mathbf k_{\mathrm{BGM}}$ and $\mathbf q_{\mathrm{ref}} =\mathbf k_{\mathrm{ref}}-\mathbf k_{\mathrm{BGM}} =\rho\,\sigma(\cos\theta,\sin\theta)$. In these coordinates, $\Lambda_{\mathbf k_{\mathrm{BGM}}+\mathbf q_{\mathrm{ref}}, \mathbf k_{\mathrm{BGM}}+\mathbf q}$ and $V_{\mathbf k_{\mathrm{BGM}}+\mathbf q_{\mathrm{ref}}, \mathbf k_{\mathrm{BGM}}+\mathbf q}$ are denoted by $\widetilde{\Lambda}_{\mathbf q_{\mathrm{ref}}\mathbf q}$ and $\widetilde V_{\mathbf q_{\mathrm{ref}}\mathbf q}$, respectively, with $\widetilde V_{\mathbf q_{\mathrm{ref}}\mathbf q} = V_0(\mathbf q_{\mathrm{ref}}-\mathbf q) \widetilde{\Lambda}_{\mathbf q_{\mathrm{ref}}\mathbf q}$. Here, $\sigma$ is the momentum-space standard deviation of the lowest-exciton probability distribution $|\varphi(\mathbf k)|^2$ about the band-gap minimum.

Figure~\ref{fig:fig3} shows the case $\theta=0$ for $\rho=0$, $0.5$, and $1.0$, while results for other directions are provided in the \supl \textrm{IV}. The red point marks $\mathbf q_{\mathrm{ref}}$, the black cross denotes $\mathbf q=\mathbf 0$, and the black circle centred at the band-gap minimum has radius $\sigma$. Figures~\ref{fig:fig3}(a--c) show $|\widetilde{\Lambda}_{\mathbf q_{\mathrm{ref}}\mathbf q}|$ over the full momentum window, whereas Figs.~\ref{fig:fig3}(d--f) magnify the region most relevant to the exciton.

At $\mathbf q=\mathbf q_{\mathrm{ref}}$, the two electron--hole product Bloch frames coincide, giving $d_{\mathrm{HS}}^{eh} (\mathbf k_{\mathrm{ref}},\mathbf k_{\mathrm{ref}})=0$ and $|\widetilde{\Lambda}_{\mathbf q_{\mathrm{ref}} \mathbf q_{\mathrm{ref}}}|=1$. As $\mathbf q$ moves away from $\mathbf q_{\mathrm{ref}}$, the finite Hilbert--Schmidt distance becomes appreciable and the overlap magnitude falls well below unity. This reduction is pronounced over momentum separations comparable to $\sigma$, with the overlap strongly suppressed across a substantial portion of the momentum region sampled by the exciton. The exciton therefore probes product Bloch frames that cannot be treated as approximately parallel over its relevant momentum range.

For $\mathbf q_{\mathrm{ref}}=\mathbf 0$, the overlap distribution is approximately symmetric around the band-gap minimum. A finite displacement of $\mathbf q_{\mathrm{ref}}$ produces an increasingly anisotropic pattern, with the overlap decreasing at different rates along different momentum directions. This anisotropy originates from the pronounced enhancement of the quantum metric near the band-gap minimum. For a displaced reference momentum, a direct path to a point on the opposite side of the minimum traverses this strongly enhanced metric region, whereas a path extending outward on the same side does not. Although this path length and the finite Hilbert--Schmidt distance need not coincide quantitatively at large separations, their directional variation provides a geometric explanation for the observed anisotropy. Accordingly, for $\mathbf q_{\mathrm{ref}}\neq\mathbf 0$, the overlap retains larger values for momenta with polar angles similar to that of $\mathbf q_{\mathrm{ref}}$, producing a distribution elongated along approximately the same radial direction from the band-gap minimum. Such momentum dependence cannot be represented by a momentum-independent renormalization of the interaction. This effect can be seen for other angles of $\mathbf q_\text{ref}$ in \supl \textrm{IV} (Fig. S1, S2, and S3). 

Figures~\ref{fig:fig3}(g--i) show $|\widetilde V_{\mathbf q_{\mathrm{ref}}\mathbf q}|/|V_0(\sigma)|$ for visual comparison, where $V_0(\sigma)$ denotes the screened Coulomb factor evaluated at a momentum-transfer magnitude $\sigma$. The momentum dependence of the projected interaction reflects both the screened Coulomb factor $V_0(\mathbf q_{\mathrm{ref}}-\mathbf q)$ and the Bloch-state overlap encoded in $\widetilde{\Lambda}_{\mathbf q_{\mathrm{ref}}\mathbf q}$.
Because $|\widetilde{\Lambda}_{\mathbf q_{\mathrm{ref}}\mathbf q}|\leq 1$,
the form factor suppresses the interaction magnitude relative to the
unity-overlap reference and introduces an additional directional
modulation near the band-gap minimum. This nonuniform suppression limits the Coulomb-induced spreading of the exciton wavefunction in momentum space, providing the mechanism underlying the enhancement of the exciton radius examined below.

\begin{figure}[!t]
\centering
\includegraphics[width=\linewidth]{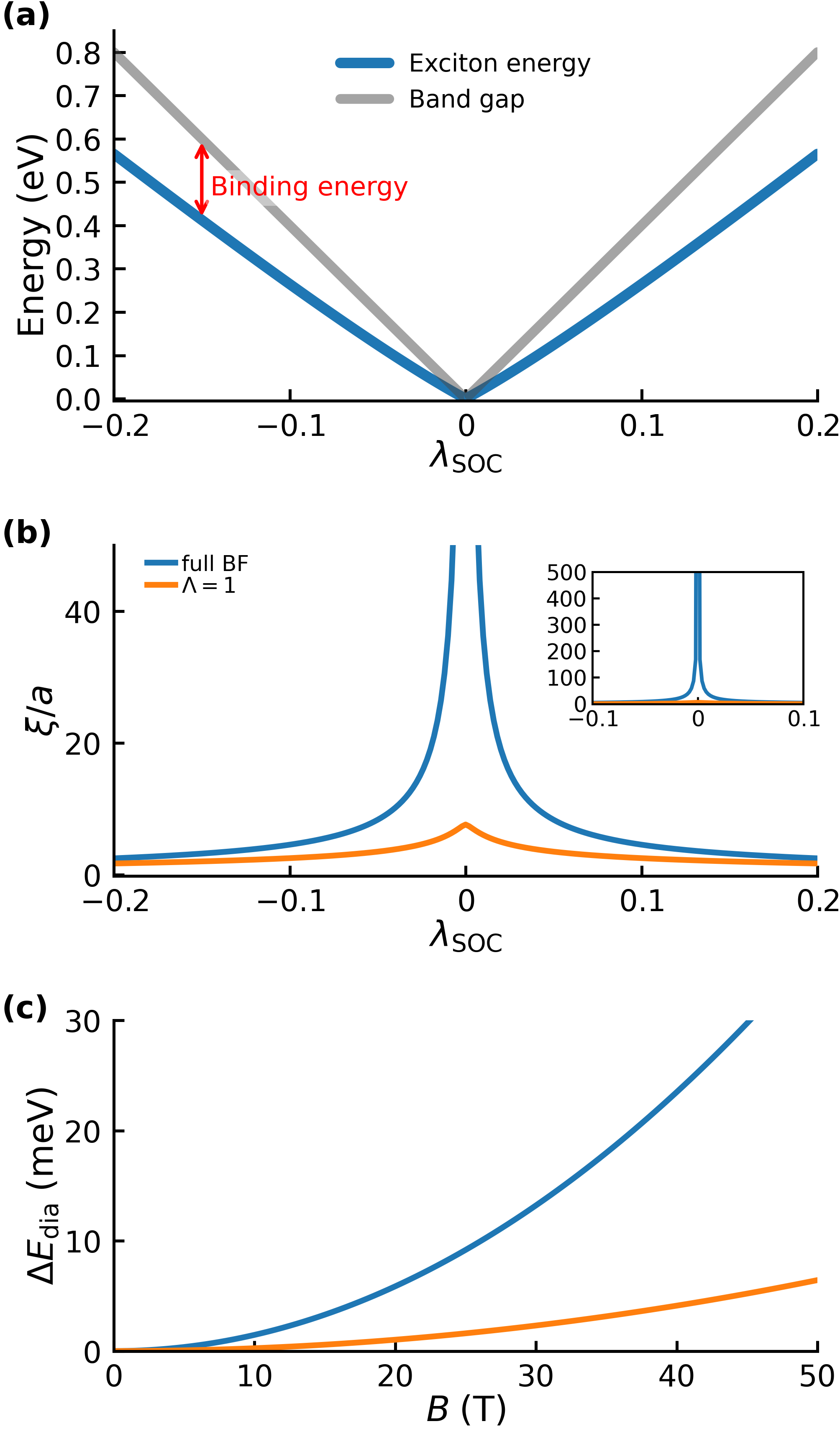}
\caption{
\textbf{Exciton-radius enhancement and diamagnetic response.}
(a) Energy of the lowest exciton state and the minimum direct band gap as functions of $\lambda_{\rm SOC}$. The difference between them is the binding energy (red arrow). (b) Corresponding radii $\xi/a$ from the Bloch-frame calculation (blue) and the unity-overlap reference (orange). The inset shows the rapid growth near the gap closing on an extended scale. (c) Quadratic diamagnetic shifts estimated from $\Delta E_{\mathrm{dia}}(B)
=e^2\xi^2B^2/(8m_{\mathrm r})$ in the small-gap regime. The unity-overlap reference is obtained by replacing $\xi$ with $\xi^{(0)}$.
}
\label{fig:fig4}
\end{figure}

\subsection*{Exciton-radius enhancement and diamagnetic response}
\label{sec:radius_diamagnetic_response}

We now examine how the quantum-distance-induced suppression of the projected interaction affects the energy and internal radius of the exciton. Figure~\ref{fig:fig4}(a) shows the exciton energy together with the minimum direct band gap as functions of $\lambda_{\rm SOC}$. The exciton remains below the electron--hole continuum for all nonzero values of $\lambda_{\rm SOC}$ considered. As $|\lambda_{\rm SOC}|$ decreases, both the direct gap and the exciton energy decrease towards the band-gap closing. Their separation also becomes progressively smaller, corresponding to a reduction in the exciton binding energy.

The corresponding internal exciton radius is shown in Fig.~\ref{fig:fig4}(b). The blue curve denotes the projected calculation retaining the momentum-dependent electron--hole Bloch frame, whereas the orange curve denotes the unity-overlap reference. In the reference calculation, the single-particle dispersions and screened Coulomb potential $V_0$ are retained, but the form factor is set to unity and the radius is evaluated using the ordinary momentum derivative. The radius obtained with the momentum-dependent Bloch frame increases sharply near the band-gap closing and becomes substantially larger than its unity-overlap reference value.

The growing separation between the two results is consistent with the interaction mechanism identified above. Near the band-gap minimum, finite quantum distance reduces the Bloch-state overlap magnitudes and thereby suppresses the projected coupling between different relative momenta compared to the unity-overlap reference. The resulting exciton wavefunction is more narrowly distributed in momentum space and therefore corresponds to a larger real-space electron--hole separation. This comparison should not be interpreted as a decomposition into independent envelope and Berry-connection contributions. The form factor changes the exciton eigenfunction itself, while the radius of the resulting state is evaluated using the covariant relative-position operator. The difference between the two radii therefore quantifies the net effect of the momentum-dependent Bloch frame within the band-projected description.

The radius enhancement becomes particularly pronounced near $\lambda_{\rm SOC}=0$, where the radius obtained with the momentum-dependent Bloch frame reaches many lattice spacings, whereas the unity-overlap reference radius increases much more moderately. This contrast shows that the closing of the single-particle gap alone does not account for the calculated enlargement. Rather, the rapid momentum variation of the electron--hole Bloch frame and the associated suppression of finite-momentum coupling substantially reconstruct the internal exciton state. No isolated-band exciton interpretation is assigned at $\lambda_{\rm SOC}=0$ itself, because the projected conduction and valence bands cease to be isolated at the gap closing. The curves therefore describe the approach to the transition from the gapped regime.

The enhanced radius has an experimentally accessible consequence in the diamagnetic response. Figure~\ref{fig:fig4}(c) shows the quadratic energy shifts estimated from the calculated radii in the small-gap regime. Because the diamagnetic coefficient scales as $\xi^2$, the calculation retaining the momentum-dependent Bloch frame yields a substantially stronger response than the unity-overlap reference. The diamagnetic response thus provides a magneto-optical signature of the quantum-distance-induced enhancement of the internal exciton radius. The plotted curves represent only the quadratic diamagnetic contribution within the weak-field effective-mass approximation. A complete finite-field spectrum would additionally require Zeeman terms, magnetic-field-induced coupling to other exciton states, and corrections beyond the effective-mass approximation.

\section*{Discussion}

Our results identify a mechanism by which momentum-space Bloch geometry controls the internal size of an exciton. Near a band-gap closing, the electron--hole product Bloch frame varies rapidly around the band-gap minimum, generating appreciable quantum distance across the momentum range sampled by the exciton. The resulting reduction of the scattering form factor suppresses the projected Coulomb coupling relative to the unity-overlap reference. Because this coupling spreads the exciton wavefunction over the BZ, its suppression narrows the momentum-space distribution and enlarges the electron--hole separation in real space.

The unity-overlap reference separates this effect from conventional changes associated with the closing gap. The calculation retaining the momentum-dependent electron--hole Bloch frame and the unity-overlap reference use the same single-particle dispersions and screened Coulomb potential $V_0(\mathbf k-\mathbf k')$, but differ in whether the electron--hole form factor and the corresponding Bloch-frame structure of the relative-position operator are retained. The growing difference between the two results therefore quantifies the total effect of the momentum-dependent Bloch frame within the projected two-band description, rather than a change caused solely by effective masses, the continuum threshold, or screening.

This mechanism is distinct from treating band geometry as an additive correction to a fixed hydrogenic exciton. Berry curvature and related geometric quantities are known to modify exciton spectra and optical properties \cite{Srivastava2015,Zhou2015}, and recent studies have linked quantum geometry to excitonic topology and transport \cite{Jankowski2025OrganicExcitons,Thompson2025ExcitonTransport}. Here, the Bloch-state overlap enters directly in the interaction kernel and reshapes the exciton eigenfunction itself. The radius is then evaluated with the covariant relative-position operator for the modified state. The dominant effect is therefore the geometry-induced reconstruction of the exciton wavefunction, rather than an independent Berry-connection correction applied afterwards.

The local quantum metric and the finite Hilbert--Schmidt distance provide complementary descriptions of this reconstruction. The metric characterizes infinitesimal Bloch-frame variation and becomes strongly enhanced near the band-gap minimum, whereas the Hilbert--Schmidt distance directly measures the finite separation between the product Bloch frames sampled by the exciton. Their relation explains the pronounced and anisotropic form-factor suppression near the gap closing. In particular, paths crossing the strongly enhanced metric region exhibit greater Bloch-frame variation than paths extending outward on the same side, producing a projected interaction that cannot be represented by a momentum-independent renormalization.

The role of topology must nevertheless be distinguished from that of quantum geometry. Although the Chern-number change is accompanied here by a gap closing and a rapid reorganization of the Bloch states, the Chern number alone does not determine the exciton radius. Strong enhancement occurs when the exciton wavefunction has substantial weight in a momentum region, where the Bloch frame varies rapidly and the form factor is substantially reduced. A topological transition is therefore a natural route to this regime, but the more general criterion is appreciable finite quantum distance within the momentum support of the exciton.

The enlarged radius produces an experimentally accessible diamagnetic signature. Within the weak-field effective-mass approximation, $\gamma_{\mathrm{dia}}=e^2\xi^2/(8m_{\mathrm r})$, so the radius enhancement yields a stronger quadratic magnetic-field shift \cite{Stier2016Diamagnetic,Goryca2019HighField}. Quantitative extraction requires a reliably determined electron--hole reduced mass and a magnetic-field range in which the quadratic approximation remains valid.

The Lieb-lattice model provides a controlled demonstration rather than a material-specific prediction. The calculation assumes isolated conduction and valence bands, zero centre-of-mass momentum, and static screening; exactly at the gap closing, the isolated-band projection is no longer valid. In realistic systems, multiband mixing, dynamical screening, disorder, phonons, and additional valleys may alter the quantitative enhancement. Nevertheless, the mechanism should extend to systems near band inversions, avoided crossings, and topological gap closings whenever the exciton samples a rapidly varying Bloch frame. Exciton size thereby becomes a sensitive two-particle manifestation of finite-momentum quantum geometry.

Looking forward, our results highlight the important role of quantum geometry in determining exciton properties and open a route to optoelectronic design that exploits the quantum structure of the constituent electron and hole states, rather than relying solely on semiclassical quantities such as effective mass and dielectric screening.

\section*{Acknowledgements}
The authors thank Bartomeu Monserrat, Wojciech J. Jankowski and Robert-Jan Slager for useful discussions. 
J.-H.C and S.-W.K acknowledge support from the research fund of Hanyang University (HY-202600000001189). J.J.P.T  acknowledges support from a EPSRC Programme grant EP/W017091/1.

\bibliographystyle{naturemag}

\section*{Author Contributions Statement} All authors discussed the results and contributed to the writing of the manuscript. The final form of the manuscript, including Methods and Supplementary Information, benefitted from input from all authors. 

\section*{Competing Interests Statement} The authors declare no competing interests.


\newpage

\newpage
\clearpage

\end{document}


\title{Supplementary Information: Enhancement of exciton radius near a band-gap closing through quantum geometry}
\author{Jin-Hyung~Choi}
\affiliation{Department of Physics, Hanyang University, Seoul 04763, Republic of Korea}
\affiliation{Department of Physics, Ajou University, Suwon 16499, Republic of Korea}

\author{Sang-Hoon~Han}
\affiliation{Department of Physics, Ajou University, Suwon 16499, Republic of Korea}

\author{Young-Kwon~Han}
\affiliation{Department of Physics, Hanyang University, Seoul 04763, Republic of Korea}

\author{Sun-Woo~Kim}
\thanks{}
\email{sunwookim@hanyang.ac.kr}
\affiliation{Department of Physics, Hanyang University, Seoul 04763, Republic of Korea}

\author{Jun-Won Rhim}
\email{jwrhim@ajou.ac.kr}
\affiliation{Department of Physics, Ajou University, Suwon 16499, Republic of Korea}

\author{Joshua J. P. Thompson}
\email{jjt56@cam.ac.uk}
\affiliation{Department of Materials Science and Metallurgy, University of Cambridge,
27 Charles Babbage Road, Cambridge CB3 0FS, United Kingdom}

\date{\today}

\date{\today}

\maketitle

\tableofcontents

\newpage

\title{}










\date{\today}

\section{Position operator}
\label{app:position_operator}

The position operator in the crystal-momentum representation was formulated by Blount in the context of band theory \cite{Blount1962}. This representation is required because the wave vector $\mathbf k$ in Bloch's theorem is a crystal momentum rather than a plane-wave momentum. A Bloch state contains both a plane-wave factor $e^{i\mathbf k\cdot\mathbf r}$ and a momentum-dependent cell-periodic part $u_{n\mathbf k}(\mathbf r)$ that defines the Bloch frame. Consequently, after projection onto a band or a restricted set of bands, a derivative with respect to crystal momentum acts on both the momentum-space coefficient and the Bloch frame. The projected position operator therefore contains an ordinary momentum derivative together with the Berry connection of the selected Bloch frame.

We use the Bloch convention
\begin{equation}
\psi_{n\mathbf k}(\mathbf r)
=
e^{i\mathbf k\cdot\mathbf r}
u_{n\mathbf k}(\mathbf r),
\label{eq:app_bloch_convention}
\end{equation}
where $\psi_{n\mathbf k}$ is the full Bloch wave function and $u_{n\mathbf k}$ is its cell-periodic part. The tight-binding orbital $j$ is located at $\mathbf R+\boldsymbol\tau_j$, where $\mathbf R$ is a Bravais-lattice vector and $\boldsymbol\tau_j$ is its intracell position.

The atomic embedding is implemented using Convention I for the tight-binding Bloch basis, in which the orbital position $\boldsymbol\tau_j$ is included in the Bloch phase \cite{Vanderbilt2018BerryPhases,YusufalyVanderbiltCoh2022PythTB}:
\begin{equation}
|\chi_{\mathbf k j}^{\rm I}\rangle
=
\sum_{\mathbf R}
e^{i\mathbf k\cdot(\mathbf R+\boldsymbol\tau_j)}
|\phi_{\mathbf Rj}\rangle .
\label{eq:app_tb_convention_I_basis}
\end{equation}
Equivalently, the full Bloch wave function evaluated at $\mathbf R+\boldsymbol\tau_j$ is
\begin{equation}
\psi_{n\mathbf k}(\mathbf R+\boldsymbol\tau_j)
=
e^{i\mathbf k\cdot(\mathbf R+\boldsymbol\tau_j)}
u_{n\mathbf k,j}.
\label{eq:app_atomic_embedding_bloch_wave}
\end{equation}
This convention differs from Convention II,
\begin{equation}
|\chi_{\mathbf k j}^{\rm II}\rangle
=
\sum_{\mathbf R}
e^{i\mathbf k\cdot\mathbf R}
|\phi_{\mathbf Rj}\rangle ,
\label{eq:app_tb_convention_II_basis}
\end{equation}
by a momentum-dependent unitary transformation in orbital space. Accordingly, the separate derivative, Berry-connection, and embedding contributions depend on the Bloch convention, whereas the physical matrix element of the position operator does not.

A state in a selected band subspace can be expanded as
\begin{equation}
\Phi(\mathbf r)
=
\int_{\rm BZ}
\frac{d^2 k}{(2\pi)^2}
\sum_n
\psi_{n\mathbf k}(\mathbf r)
\varphi_n(\mathbf k).
\label{eq:app_general_bloch_state}
\end{equation}
Using
$\hat r_\mu e^{i\mathbf k\cdot\mathbf r}
=-i\partial_\mu e^{i\mathbf k\cdot\mathbf r}$,
where $\partial_\mu\equiv\partial/\partial k_\mu$, gives
\begin{align}
\hat r_\mu
\psi_{n\mathbf k}(\mathbf r)\varphi_n(\mathbf k)
={}&
-i\partial_\mu
\left[
\psi_{n\mathbf k}(\mathbf r)\varphi_n(\mathbf k)
\right]
\nonumber\\
&+
e^{i\mathbf k\cdot\mathbf r}
i\partial_\mu u_{n\mathbf k}(\mathbf r)
\varphi_n(\mathbf k)
+
\psi_{n\mathbf k}(\mathbf r)
i\partial_\mu\varphi_n(\mathbf k).
\label{eq:app_position_action_decomposed}
\end{align}

The first term in Eq.~\eqref{eq:app_position_action_decomposed} is a total derivative over the chosen fundamental Brillouin-zone domain:
\begin{equation}
\int_{\rm FBZ}
d^2 k\,
\partial_\mu F_n(\mathbf k,\mathbf r)
=
\int_{\partial{\rm FBZ}}
d S\,
n_\mu(\mathbf k)
F_n(\mathbf k,\mathbf r),
\label{eq:app_boundary_divergence}
\end{equation}
where
$F_n(\mathbf k,\mathbf r)
=\psi_{n\mathbf k}(\mathbf r)\varphi_n(\mathbf k)$.
For a piecewise-smooth Bloch frame satisfying the appropriate sewing conditions, the values on paired boundary faces are identified by
\begin{equation}
F_n(\mathbf k+\mathbf G,\mathbf r)
=
F_n(\mathbf k,\mathbf r),
\label{eq:app_F_sewing}
\end{equation}
where $\mathbf G$ is a reciprocal-lattice vector. Since the outward normals on paired faces have opposite signs, their boundary contributions cancel:
\begin{equation}
\int_{\partial{\rm FBZ}}
d S\,
n_\mu(\mathbf k)
F_n(\mathbf k,\mathbf r)
=
0.
\label{eq:app_boundary_cancellation}
\end{equation}

Using the orthonormality of the full Bloch states, the remaining terms
give
\begin{equation}
\langle\Phi_1|\hat r_\mu|\Phi_2\rangle
=
\int_{\rm BZ}
\frac{d^2 k}{(2\pi)^2}
\sum_{mn}
\varphi_{1,m}^*(\mathbf k)
\left[
\mathcal A^\mu_{mn}(\mathbf k)
+
i\delta_{mn}\partial_\mu
\right]
\varphi_{2,n}(\mathbf k),
\label{eq:app_position_matrix_element}
\end{equation}
where
\begin{equation}
\mathcal A^\mu_{mn}(\mathbf k)
=
i
\langle u_{m\mathbf k}|
\partial_\mu u_{n\mathbf k}\rangle
\label{eq:app_connection_matrix}
\end{equation}
is the Berry-connection matrix of the selected Bloch frame. Equation \eqref{eq:app_position_matrix_element} explicitly shows that the projected position operator contains both the ordinary derivative acting on the momentum-space coefficient and the Berry connection arising from the momentum dependence of the basis \cite{Blount1962,Ahn2022RiemannianOptics}.

In Convention I, the reciprocal-space boundary condition is represented by a sewing matrix in orbital space. In a periodic representation of the full Bloch states,
\begin{equation}
\psi_{n,\mathbf k+\mathbf G}(\mathbf r)
=
\psi_{n\mathbf k}(\mathbf r).
\label{eq:app_full_bloch_periodicity}
\end{equation}
Using Eq.~\eqref{eq:app_bloch_convention} then gives
\begin{equation}
e^{i\mathbf G\cdot\mathbf r}
u_{n,\mathbf k+\mathbf G}(\mathbf r)
=
u_{n\mathbf k}(\mathbf r).
\label{eq:app_cell_periodic_sewing_continuum}
\end{equation}
At the orbital position $\mathbf r=\mathbf R+\boldsymbol\tau_j$, this becomes
\begin{equation}
u_{n,\mathbf k+\mathbf G,j}
=
e^{-i\mathbf G\cdot\boldsymbol\tau_j}
u_{n\mathbf k,j}.
\label{eq:app_component_sewing}
\end{equation}
The tight-binding eigenvectors therefore obey
\begin{equation}
u_{n,\mathbf k+\mathbf G}
=
S_{\mathbf G}u_{n\mathbf k},
\qquad
[S_{\mathbf G}]_{jj'}
=
e^{-i\mathbf G\cdot\boldsymbol\tau_j}
\delta_{jj'}.
\label{eq:app_sewing_matrix}
\end{equation}
This sewing matrix follows directly from the atomic embedding in Convention I and is included in the evaluation of geometric quantities and projected position-operator matrix elements.

For the electron--hole problem considered in the main text, the relevant product Bloch frame is
\begin{equation}
|U_{\mathbf k}\rangle
=
|u_{c\mathbf k}\rangle
\otimes
|u^*_{v\mathbf k}\rangle .
\label{eq:app_eh_product_frame}
\end{equation}
Its Berry connection is
\begin{equation}
\mathcal A^{eh}_\mu(\mathbf k)
=
i\langle U_{\mathbf k}|
\partial_\mu U_{\mathbf k}\rangle
=
\mathcal A^c_\mu(\mathbf k)
-
\mathcal A^v_\mu(\mathbf k).
\label{eq:app_eh_connection}
\end{equation}
The projected electron--hole relative-position operator is consequently
\begin{equation}
\hat r_\mu
=
iD_\mu,
\qquad
D_\mu
=
\partial_\mu-i\mathcal A^{eh}_\mu .
\label{eq:app_eh_relative_position}
\end{equation}

The covariant derivative acts on the exciton section formed jointly by the coefficient $\varphi_\eta(\mathbf k)$ and the electron--hole product frame. Under a local change of Bloch frame, these two components transform oppositely. The covariant derivative of the section is therefore independent of the chosen local representation and is represented by $D_\mu\varphi_\eta$ in the selected frame. In the unity-overlap reference, the momentum dependence of the product Bloch frame is removed consistently from both the interaction and the radius evaluation, so that $D_\mu$ is replaced by the ordinary derivative $\partial_\mu$.

\section{Topology of the spin--orbit-coupled Lieb-lattice bands}
\label{app:lieb_topology}

Here we summarize the band topology of the spin--orbit-coupled Lieb-lattice model used in the main text. In the three-sublattice basis, the Bloch Hamiltonian is
\begin{equation}
H(\mathbf k)
=
\begin{pmatrix}
0
&
2\cos(k_y/2)
&
-4i\lambda_{\rm SOC}\sin(k_x/2)\sin(k_y/2)
\\
2\cos(k_y/2)
&
0
&
2\cos(k_x/2)
\\
4i\lambda_{\rm SOC}\sin(k_x/2)\sin(k_y/2)
&
2\cos(k_x/2)
&
0
\end{pmatrix}.
\label{eq:lieb_hamiltonian_appendix}
\end{equation}
The first, second, and third components correspond to the three sublattice orbitals in the Lieb-lattice unit cell. The spin--orbit term opens a gap at the band-touching points of the spinless Lieb lattice and produces three isolated bands for nonzero $\lambda_{\rm SOC}$, with the lower and upper bands carrying opposite Chern numbers.

Because the orbitals occupy different intracell positions, the Hamiltonian is periodic in momentum space up to the sublattice sewing matrices. For reciprocal translations by $2\pi$ in the $k_x$ and $k_y$ directions,
\begin{equation}
H(k_x+2\pi,k_y)
=
S_xH(k_x,k_y)S_x^\dagger,
\qquad
H(k_x,k_y+2\pi)
=
S_yH(k_x,k_y)S_y^\dagger,
\label{eq:sewing_hamiltonian_appendix}
\end{equation}
where
\begin{equation}
S_x
=
\begin{pmatrix}
1&0&0\\
0&1&0\\
0&0&-1
\end{pmatrix},
\qquad
S_y
=
\begin{pmatrix}
-1&0&0\\
0&1&0\\
0&0&1
\end{pmatrix}.
\label{eq:sewing_matrices_appendix}
\end{equation}
These matrices are included when evaluating overlaps and geometric quantities across a Brillouin-zone boundary.

For band $n$, the Berry connection is
\begin{equation}
\mathcal A_{n,\mu}(\mathbf k)
=
i
\langle u_{n\mathbf k}|
\partial_\mu u_{n\mathbf k}\rangle ,
\label{eq:berry_connection_appendix}
\end{equation}
and the corresponding Berry curvature is
\begin{equation}
\mathcal F_n(\mathbf k)
=
\partial_x\mathcal A_{n,y}(\mathbf k)
-
\partial_y\mathcal A_{n,x}(\mathbf k).
\label{eq:berry_curvature_appendix}
\end{equation}
The Chern number is
\begin{equation}
C_n
=
\frac{1}{2\pi}
\int_{\rm BZ}
d^2 k\,
\mathcal F_n(\mathbf k).
\label{eq:chern_number_appendix}
\end{equation}

Ordering the bands from lowest to highest energy gives
\begin{equation}
(C_1,C_2,C_3)
=
(+1,0,-1),
\qquad
\lambda_{\rm SOC}>0,
\label{eq:chern_positive_appendix}
\end{equation}
and
\begin{equation}
(C_1,C_2,C_3)
=
(-1,0,+1),
\qquad
\lambda_{\rm SOC}<0.
\label{eq:chern_negative_appendix}
\end{equation}
Changing the sign of $\lambda_{\rm SOC}$ therefore reverses the chirality of the lower and upper Chern bands, whereas varying $|\lambda_{\rm SOC}|$ at fixed sign preserves their Chern numbers as long as the gaps remain open. The middle band is Chern-trivial, while the lower and upper bands carry opposite Chern numbers.

The excitons considered in the main text are formed between the upper conduction band and the middle valence band. The corresponding electron--hole product-frame connection is
\begin{equation}
\mathcal A^{eh}_\mu
=
\mathcal A^c_\mu
-
\mathcal A^v_\mu,
\label{eq:eh_connection_appendix}
\end{equation}
and its curvature is
\begin{equation}
\mathcal F^{eh}
=
\mathcal F_c
-
\mathcal F_v.
\label{eq:eh_curvature_appendix}
\end{equation}
Since the middle band has zero Chern number, the integrated electron--hole curvature is determined by the Chern number of the upper band. For $\lambda_{\rm SOC}>0$, the upper band has $C_c=-1$, whereas for $\lambda_{\rm SOC}<0$, it has $C_c=+1$.

The Chern number characterizes the global topology of the isolated bands, whereas the local Bloch-state geometry determines the momentum-dependent overlaps and Berry connection entering the projected interaction and relative-position operator. The Chern numbers remain fixed within each gapped phase, while these local geometric quantities
change strongly as the band-gap closing is approached.

\section{Non-uniform discretization of the Wannier equation}
\label{app:nonuniform_wannier}

The numerical solution of the momentum-space Wannier equation on a finite open-boundary domain must satisfy two requirements. First, the domain must be sufficiently large that the exciton wave function is negligible near its boundary. Otherwise, the open boundary can artificially distort $\varphi_\eta(\mathbf k)$. Second, the covariant
derivative $D_\mu\varphi_\eta$ used in the full radius calculation, or the corresponding ordinary derivative in the unity-overlap reference, must be resolved accurately over the momentum region in which the exciton wave function has appreciable weight.

The derivative-sensitive region is considerably smaller than the momentum-space buffer required to suppress boundary effects. A uniform mesh would therefore impose an unnecessarily fine resolution throughout the full domain. We instead use a non-uniform polar finite-volume mesh centred at the band-gap minimum. The radial spacing increases with the distance from the mesh centre, providing fine resolution in the region relevant to the exciton and progressively coarser resolution in the outer buffer.

For each value of $\lambda_{\rm SOC}$, the zero-centre-of-mass band-projected Wannier equation is
\begin{equation}
\Delta(\mathbf k)\varphi_\eta(\mathbf k)
+
\int_{\mathcal D}
\frac{d^2 k'}{a_0^2(2\pi)^2}\,
V_{\mathbf k\mathbf k'}
\varphi_\eta(\mathbf k')
=
E_\eta\varphi_\eta(\mathbf k),
\label{eq:app_nonuniform_wannier_continuum}
\end{equation}
where
\begin{equation}
\Delta(\mathbf k)
=
\varepsilon_c(\mathbf k)-\varepsilon_v(\mathbf k)
\label{eq:app_nonuniform_direct_gap}
\end{equation}
and
\begin{equation}
V_{\mathbf k\mathbf k'}
=
V_0(\mathbf k-\mathbf k')
\Lambda_{\mathbf k\mathbf k'}.
\label{eq:app_nonuniform_projected_interaction}
\end{equation}
The electron--hole form factor is
\begin{equation}
\Lambda_{\mathbf k\mathbf k'}
=
\langle u_{c\mathbf k}|u_{c\mathbf k'}\rangle
\langle u_{v\mathbf k'}|u_{v\mathbf k}\rangle .
\label{eq:app_nonuniform_form_factor}
\end{equation}
The unity-overlap reference is obtained by setting $\Lambda_{\mathbf k\mathbf k'}=1$ while retaining the same
single-particle dispersions and screened interaction $V_0$.

The screened attractive interaction is
\begin{equation}
V_0(q)
=
-
\frac{e^2}{2\epsilon_0\epsilon_s}
\frac{1}{q(1+r_0q)},
\qquad
r_0
=
\frac{L_w\epsilon_w}{2\epsilon_s},
\label{eq:app_nonuniform_rk_potential}
\end{equation}
where
\begin{equation}
q
=
\frac{|\mathbf k-\mathbf k'|}{a_0}.
\label{eq:app_nonuniform_physical_transfer}
\end{equation}
The momenta entering the tight-binding Hamiltonian are dimensionless, and $a_0$ is the lattice constant. In the numerical implementation, the zero-momentum-transfer component is omitted by setting the $q=0$ matrix element to zero. The same prescription is used in the full and unity-overlap calculations.

The polar mesh is centred at $\mathbf k_{\mathrm{BGM}}=(\pi,\pi)$. Its radial cells are generated with a geometrically increasing spacing, while the angular coordinate is uniformly discretized. The increasing radial cell width provides fine resolution near the band-gap minimum and a large outer buffer without requiring the same resolution throughout the full domain.

For a radial cell bounded by $\rho_l$ and $\rho_{l+1}$ and an angular cell bounded by $\theta_m$ and $\theta_{m+1}$, the representative point is placed at the angular midpoint and at the area-weighted radial average
\begin{equation}
\rho_l^{\rm mid}
=
\frac{2}{3}
\frac{\rho_{l+1}^3-\rho_l^3}
{\rho_{l+1}^2-\rho_l^2}.
\label{eq:app_nonuniform_radial_midpoint}
\end{equation}
The corresponding cell area in dimensionless crystal momentum is
\begin{equation}
A_{lm}
=
\frac{1}{2}
\left(
\rho_{l+1}^2-\rho_l^2
\right)
\left(
\theta_{m+1}-\theta_m
\right),
\label{eq:app_nonuniform_cell_area}
\end{equation}
and the quadrature weight is
\begin{equation}
W_{lm}
=
\frac{A_{lm}}{a_0^2(2\pi)^2}.
\label{eq:app_nonuniform_weight}
\end{equation}
The innermost radial cell is represented by a single point at $\mathbf k_{\mathrm{BGM}}$ carrying the full angular weight of the central disk.

Finite-volume quadrature converts Eq.~\eqref{eq:app_nonuniform_wannier_continuum} into
\begin{equation}
\Delta_i\varphi_{\eta,i}
+
\sum_j
W_jV_{ij}\varphi_{\eta,j}
=
E_\eta\varphi_{\eta,i},
\label{eq:app_nonuniform_discrete_wannier}
\end{equation}
where
\begin{equation}
\Delta_i
=
\Delta(\mathbf k_i),
\qquad
V_{ij}
=
V_0(\mathbf k_i-\mathbf k_j)\Lambda_{ij},
\qquad
\Lambda_{ij}
=
\Lambda_{\mathbf k_i\mathbf k_j}.
\label{eq:app_nonuniform_discrete_interaction}
\end{equation}
For the unity-overlap reference, $\Lambda_{ij}$ is replaced by unity.

Equation~\eqref{eq:app_nonuniform_discrete_wannier} is Hermitian with respect to the weighted inner product
\begin{equation}
\langle\phi|\psi\rangle_W
=
\sum_i
W_i\phi_i^*\psi_i.
\label{eq:app_nonuniform_weighted_inner_product}
\end{equation}
For numerical diagonalization, we define
\begin{equation}
g_{\eta,i}
=
\sqrt{W_i}\,
\varphi_{\eta,i}
\label{eq:app_nonuniform_weighted_vector}
\end{equation}
and solve the ordinary Hermitian eigenvalue problem
\begin{equation}
\sum_j
H^{(W)}_{ij}g_{\eta,j}
=
E_\eta g_{\eta,i},
\label{eq:app_nonuniform_hermitian_problem}
\end{equation}
with
\begin{equation}
H^{(W)}_{ij}
=
\Delta_i\delta_{ij}
+
\sqrt{W_iW_j}\,
V_{ij}.
\label{eq:app_nonuniform_weighted_matrix}
\end{equation}
After diagonalization, the local exciton coefficient is recovered as
\begin{equation}
\varphi_{\eta,i}
=
\frac{g_{\eta,i}}{\sqrt{W_i}},
\qquad
\sum_i
W_i
|\varphi_{\eta,i}|^2
=
1.
\label{eq:app_nonuniform_phi_recovery}
\end{equation}

The Bloch eigenvectors entering $\Lambda_{ij}$ are obtained by diagonalizing the Lieb-lattice Hamiltonian on the same non-uniform mesh. The numerical domain is treated as a contractible local momentum patch around the band-gap minimum, with the sewing matrices applied when a mesh link crosses a conventional Brillouin-zone boundary. An overlap-based phase alignment referenced to the mesh centre is used to improve the numerical smoothness of the local Bloch frame.

Under local phase changes of the conduction- and valence-band eigenvectors, $\Lambda_{ij}$ and the exciton coefficients transform covariantly, leaving the Wannier spectrum and physical observables unchanged. The finite momentum domain is chosen sufficiently large that the low-lying exciton wave functions are negligible at its open
boundary. At the exact gap-closing point, the conduction and valence bands are not isolated, and the corresponding band-projected Wannier problem and covariant radius are therefore not assigned a physical value.

\section{Angular dependence of the form-factor suppression}
\label{app:angular_dependence}

To examine the directional dependence of Fig.~3 in the main text, we repeat the calculation for reference momenta
$\mathbf q_{\mathrm{ref}} =s\sigma(\cos\theta,\sin\theta)$ at additional values of $\theta$. Supplementary
Fig.~\ref{fig:figs1-2}, and \ref{fig:figs1-3} shows $|\widetilde{\Lambda}_{\mathbf q_{\mathrm{ref}}\mathbf q}|$
and $|\widetilde V_{\mathbf q_{\mathrm{ref}}\mathbf q}|/|V_0(\sigma)|$. The pronounced suppression across the band-gap minimum and the preferential retention of coupling along the radial direction of $\mathbf q_{\mathrm{ref}}$ persist for all angles considered, while the detailed anisotropy reflects the underlying lattice symmetry. The mechanism identified in the main text is therefore not specific to $\theta=0$.

\begin{figure*}[!t]
\centering
\includegraphics[width=\textwidth]{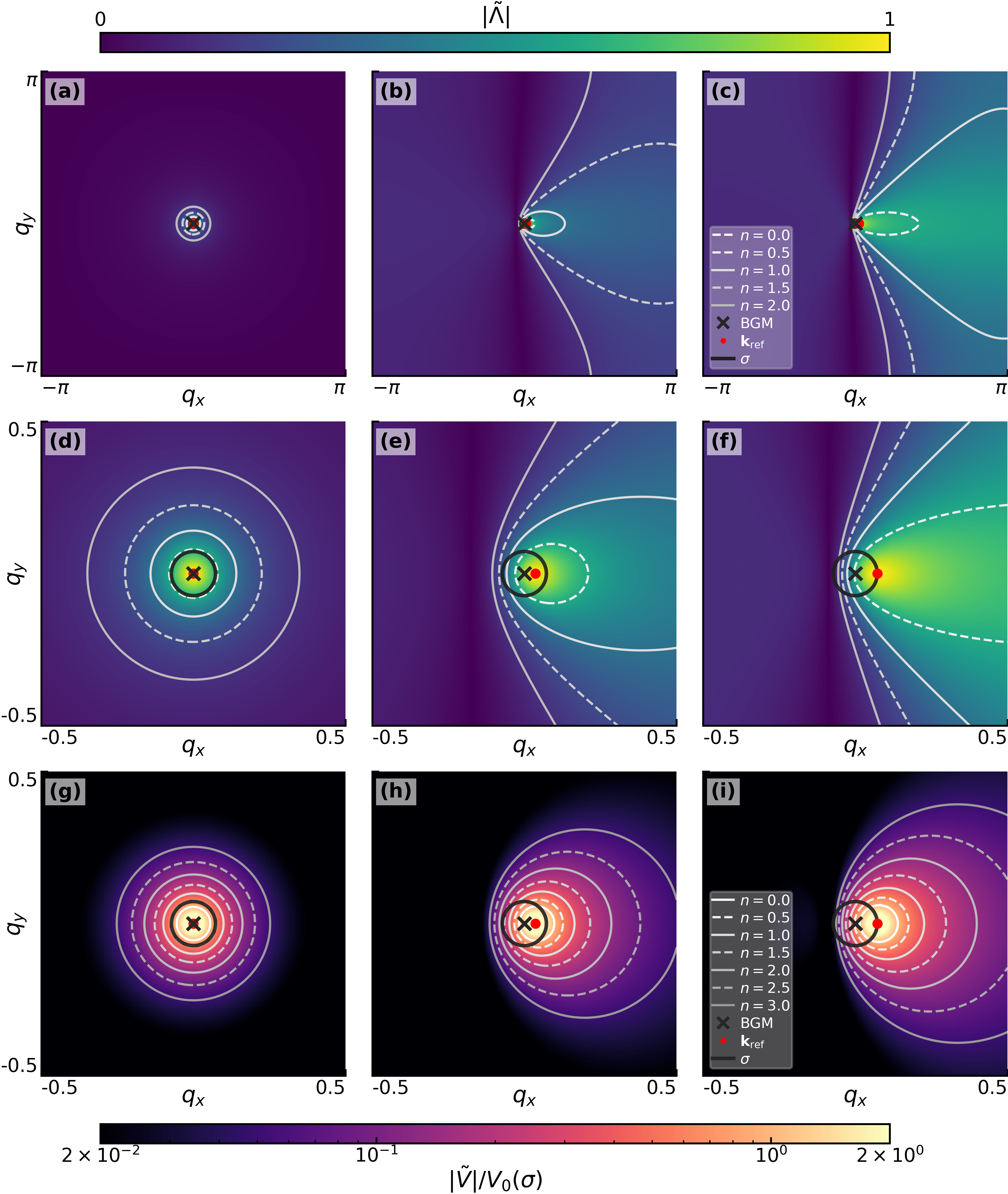}
\caption{
\textbf{Form-factor suppression for $\theta=0$.}
Magnitude of $|\widetilde{\Lambda}_{\mathbf q_{\rm ref}\mathbf q}|$ for $\mathbf q_{\rm ref}=s\sigma(\cos\theta,\sin\theta)$ with $\theta=0$. Panels (a)--(c) show the full momentum window for $s=0$, $0.5$, and $1.0$, respectively, and panels (d)--(f) show the corresponding regions around the band-gap minimum. The solid, dashed, and dotted contours mark $|\widetilde{\Lambda}|=0.5$, $0.25$, and $0.125$. The cross marks the band-gap minimum, the dashed circle has radius $\sigma$, and the open circle marks $\mathbf q_{\rm ref}$.
}
\label{fig:figs1-1}
\end{figure*}

\begin{figure*}[!t]
\centering
\includegraphics[width=\textwidth]{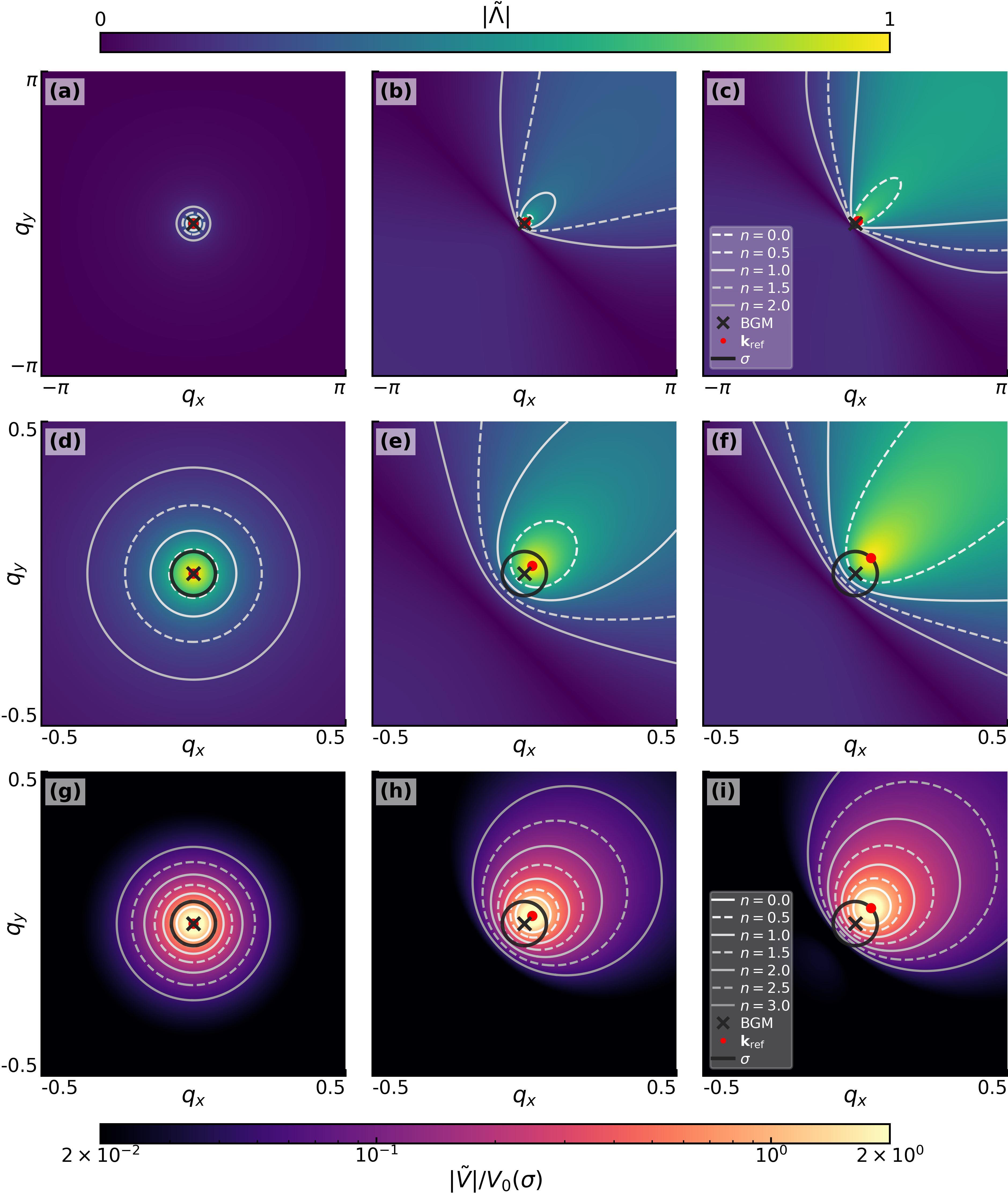}
\caption{
\textbf{Form-factor suppression for $\theta=\pi/4$.}
Magnitude of
$|\widetilde{\Lambda}_{\mathbf q_{\rm ref}\mathbf q}|$
for
$\mathbf q_{\rm ref}=s\sigma(\cos\theta,\sin\theta)$
with $\theta=\pi/4$.
Panels (a)--(c) show the full momentum window for $s=0$, $0.5$, and $1.0$, respectively, and panels (d)--(f) show the corresponding regions around the band-gap minimum. The solid, dashed, and dotted contours mark
$|\widetilde{\Lambda}|=0.5$, $0.25$, and $0.125$. The cross marks the band-gap minimum, the dashed circle has radius $\sigma$, and the open circle marks $\mathbf q_{\rm ref}$.
}
\label{fig:figs1-2}
\end{figure*}

\begin{figure*}[!t]
\centering
\includegraphics[width=\textwidth]{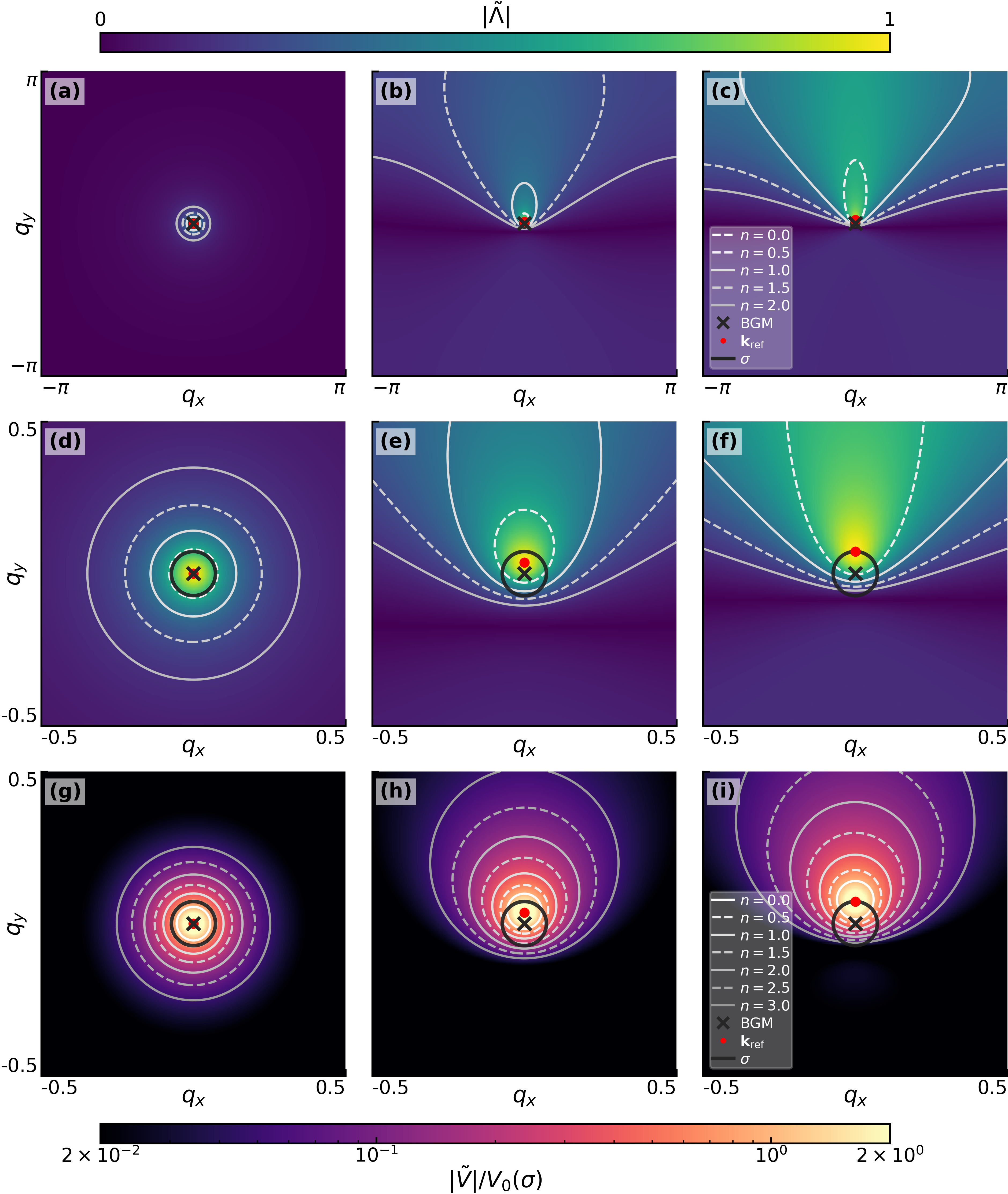}
\caption{
\textbf{Form-factor suppression for $\theta=\pi/2$.}
Magnitude of $|\widetilde{\Lambda}_{\mathbf q_{\rm ref}\mathbf q}|$ for $\mathbf q_{\rm ref}=s\sigma(\cos\theta,\sin\theta)$ with $\theta=\pi/2$. Panels (a)--(c) show the full momentum window for $s=0$, $0.5$, and $1.0$, respectively, and panels (d)--(f) show the corresponding regions around the band-gap minimum. The solid, dashed, and dotted contours mark $|\widetilde{\Lambda}|=0.5$, $0.25$, and $0.125$. The cross marks the band-gap minimum, the dashed circle has radius $\sigma$, and the open circle marks $\mathbf q_{\rm ref}$.
}
\label{fig:figs1-3}
\end{figure*}

\newpage